\documentstyle[sprocl,epsfig]{article}

\newcommand{\dr}[1]{_{\rm #1}}

\def\be{\begin{equation}}
\def\ee{\end{equation}}
\def\bea{\begin{eqnarray}}
\def\eea{\end{eqnarray}}

\begin{document}

\title{SURFACE PROPERTIES IN A RELATIVISTIC MEAN FIELD APPROACH WITH
       VECTOR-VECTOR AND SCALAR-VECTOR SELF-INTERACTIONS.}

\author{\underline{M. DEL ESTAL}, M. CENTELLES, X. VI\~NAS}

\address{Departament d'Estructura i Constituents de la Mat\`eria,
         Universitat de Barcelona, Diagonal 647, E-08028 Barcelona, Spain}


\maketitle\abstracts{ Nuclear matter and surface properties are
carefully studied for a model based on an effective hadronic lagrangian
with vector-vector and scalar-vector self-interactions. The results of
the model are compared with those of the successful
standard non-linear $\sigma-\omega$ model.}

\section{The model.}

 Quantum Hadrodynamics (QHD) and
 the relativistic treatment of nuclear systems has been a subject of
growing interest during recent years.
The $\sigma-\omega$ model of Walecka \cite{Serot} plus extensions
have been widely used to this end.      
Traditionally, the hadronic lagrangian was required to be
renormalizable.
To parametrize the density dependence of the
interaction,
cubic and quartic self-interactions of the
scalar field                                    \cite{Boguta} were introduced.
At the mean field (Hartree) level, the model has been very successful in
describing
             many properties of the atomic nucleus.


   But recently it has been proposed to look at QHD from a different
perspective, as an effective field theory \cite{Furnstahl}.               The lagrangian
has to be generalized including all the terms, {\it  non-renormalizable}
in general, that are consistent with the symmetries of the underlying
theory, QCD.
In principle, this lagrangian has infinite terms. Therefore, it is necessary
                                                                to find
suitable expansion parameters and a scheme of truncation of the generated
series. In the nuclear matter problem the meson fields divided by the
nucleon mass and their gradients are small and can be used to this
end \cite{Serot,Furnstahl}.  The guide  in this expansion should be the
concept of naturalness: it          means that all the unknown couplings
of the theory when          written in appropriate dimensionless form
should be of order unity.
                   In Ref.\ \cite{Furnstahl} the free coefficients were fitted to
several nuclear properties and it was found that the best  fits to nuclei
were indeed natural and             that it is enough to go to fourth
order in the expansion.

   In this work we will study the influence of these new couplings on
the surface properties of nuclei.
 We will work in the
semi-infinite nuclear matter geometry which is more convenient to study
    surface effects.

\section{The Energy Functional}
 The energy functional                 in     mean field approximation
    for semi-infinite nuclear matter reads

\vspace{-3mm}
\bea
{\cal E}(z) & = &  \sum_{i}
\varphi_{i}^{\dag}(z) \left \{
-i \mbox{\boldmath $\alpha$ }\cdot \mbox{\boldmath $\nabla$} + \beta \left[ M - \Phi(z)
\right]
+ W(z)
\right \} \varphi_{i}(z)
\nonumber \\[1mm]
& & + \frac{1}{2g\dr{s}^2}\left( 1 +\alpha_1\frac{\Phi(z)}{M}\right)
\left(
\mbox{\boldmath $\nabla$}\Phi(z)\right)^2
+ \left ( \frac{1}{2}
+ \frac{\kappa_3}{3!}\frac{\Phi(z)}{M}
+ \frac{\kappa_4}{4!}\frac{\Phi^2(z)}{M^2}\right ) \frac{m\dr{s}^2}{g\dr{s}^2}
\Phi^2(z)
\nonumber \\[1mm]
& & -\frac{1}{2g\dr{v}^2}\left( 1 +\alpha_2\frac{\Phi(z)}{M}\right) \left(
\mbox{\boldmath $\nabla$} W(z)  \right)^2  
- \frac{\zeta_0}{4!} \frac{1}{ g\dr{v}^2 } W^4 (z)
\nonumber \\[1mm]
& &  - \frac{1}{2}\left(1 + \eta_1 \frac{\Phi(z)}{M} + \frac{\eta_2}{2}
\frac{\Phi^2 (z)}{M^2}
\right)
 \frac{m\dr{v}^2}{g\dr{v}^2} W^2 (z)  \,,
\label{eq1}
\eea
where the index $i$ runs over all occupied states of the positive energy
spectrum,
$\Phi \equiv g\dr{s} \phi_0$ and  $ W \equiv g\dr{v} V_0$ (notation as in Ref.\
 \cite{Furnstahl}). Except for the terms with $\alpha_1$ and $\alpha_2$, the
functional (\ref{eq1}) is of fourth-order in the expansion. We retain the
fifth-order terms $\alpha_1$ and $\alpha_2$ because in Refs.\  \cite{Serot}   and
\cite{Furnstahl} they have been estimated to be numerically important at the
surface.

   Though we have studied several  surface properties such as the density
profile, the central potential, the spin-orbit potential, etc, here
              we will focus on the surface energy coefficient $E\dr{s}$ Ref.\ \cite{Myers}
                     and on the surface thickness $t$
(90\%--10\% fall-off distance of the density profile).

\vspace{-0.25cm}
\section{Results.}
\subsection{Volume terms: $\zeta_0$, $\eta_1$, $\eta_2$.}

     Figure 1a shows          the change of $E\dr{s}$ and $t$ against the
parameter $\eta_0$ that is related with the quartic vector self-interaction
coefficient $\zeta_0$ through  $\eta_0 =\sqrt{6m\dr{v}^6/(  g\dr{v}^4 \rho_0^2 \zeta_0)}$
(Ref.\ \cite{Bodmer}),                  for the        saturation properties
 $\rho_0$= 0.152 fm$^{-3}$, $a\dr{v}$= $-$16.42 MeV, $K$= 200 MeV and
 $M^{*}_{\infty}/M$ = 0.6 or 0.7, with $m\dr{s}$ = 490 MeV.  We can see that the overall effect is small
 for the natural zone (that roughly is \cite{Estal}  $2 \leq \eta_0 \leq \infty$).
 Nevertheless, there is a change of tendency with
  $M^{*}_{\infty}/M$.
 For $M^{*}_{\infty}/M$ = 0.6, both $E\dr{s}$ and $t$ decrease   while for
  $M^{*}_{\infty}/M$ = 0.7, $E\dr{s}$ increases and  $t$       decreases. This
 effect, that can also be achieved changing $K$, can help to find parametrizations
 with both $E\dr{s}$ and $t$ lying in the empirical region \cite{Estal}.

     The effect of the mixed scalar-vector self-interactions can be seen in
 Figure 1b. The figure depicts the contours of constant  $E\dr{s}$ (solid lines) and $t$
 (dashed lines)  in the plane $\eta_1-\eta_2$ for the same saturation properties
 as before (with  $M^{*}_{\infty}/M$ = 0.6).
                          On the one hand, the verticality of the lines indicates that
 the influence of the $\eta_2$ coupling                                   is
 negligible compared to the coupling      $\eta_1$. On the other hand, the
 slope   of the lines of constant $E\dr{s}$ and $t$ is very similar. Therefore,
 it is not possible to change  $t$  keeping a constant  $E\dr{s}$ (or viceversa)
 from the interplay of the $\eta_1$ and $\eta_2$ coefficients. We also have
 checked that these global trends are the same with different values of $K$ and
   $M^{*}_{\infty}/M$.

 \vspace{-0.50cm}
 \begin{figure}[h!]
   \caption{(a). $E\dr{s}$(solid line) and $t$ (dashed line)  against $\eta_0$ for
  the saturation properties indicated in the text. (b) Contours           of constant
  $E\dr{s}$ (solid) and $t$ (dashed) in the  $\eta_1-\eta_2$ plane.}
         \setlength{\unitlength}{1mm}
         \begin{picture}(100,100)
         \put(19, 15 ){\epsfxsize=7cm \epsfbox{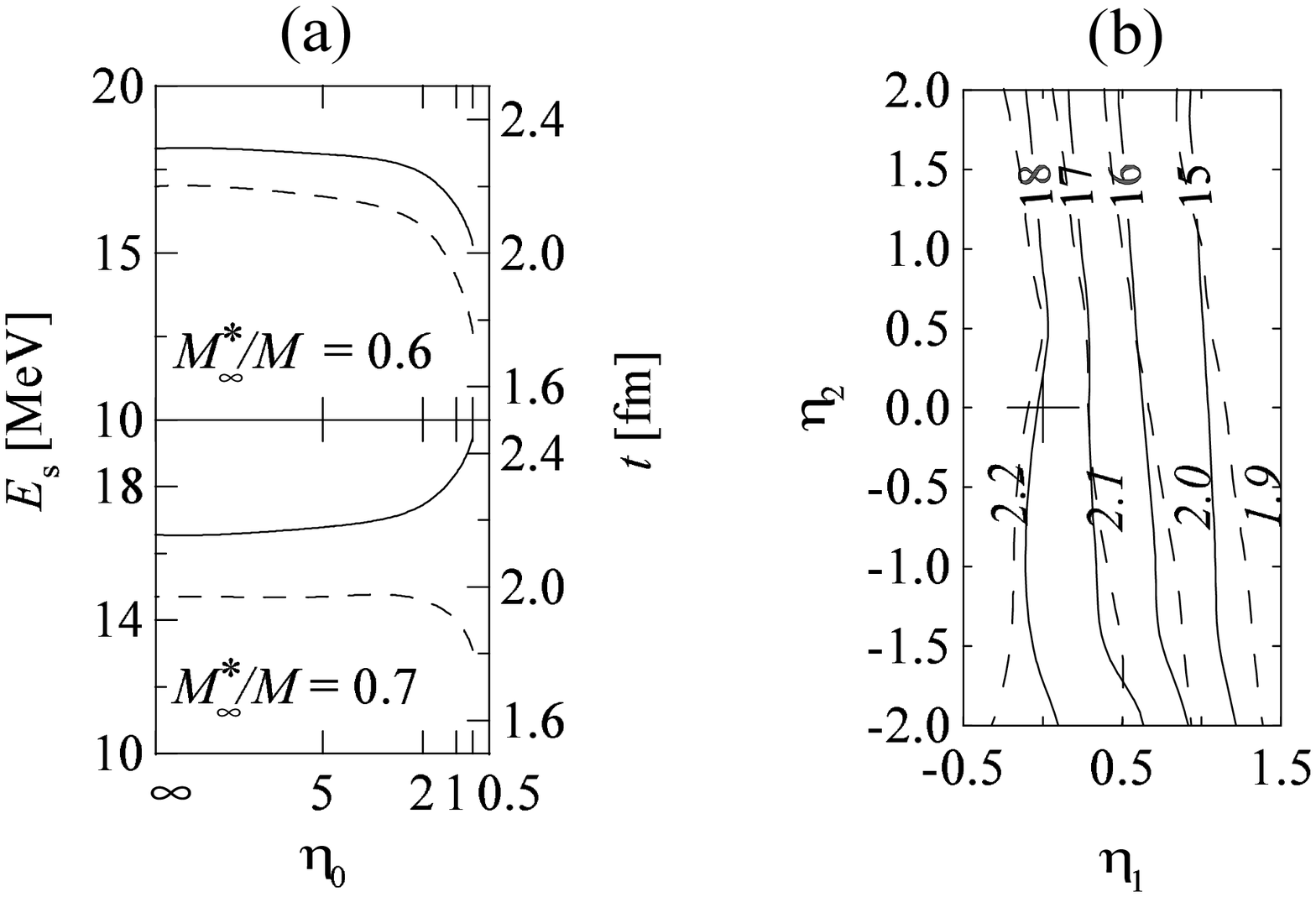}}
         \end{picture}
         \end{figure}
 \vspace{- 4.85cm}
  \subsection{Gradient terms : $\alpha_1$, $\alpha_2$.}
   Figure 2a  shows the lines of constant  $E\dr{s}$ and $t$ in the plane
$\alpha_1-\alpha_2$, for the same nuclear matter properties as before. Now, the
slopes   of the contours of constant $E\dr{s}$ and $t$ are slightly different.
Then,                 keeping a constant $E\dr{s}$ it is possible to change $t$ in a small but
appreciable margin. Also, it can be seen that the range of variation of both
$E\dr{s}$ and $t$ is wider than for the volume terms. This justifies the
inclusion in the energy functional of the gradient terms.     Figure 2b displays
     the influence on the spin-orbit strengh $V\dr{so}(z)$ (Ref.\ \cite{Hofer})   of these terms. The effect is
not negligible, at least for  $M^{*}_{\infty}/M$  = 0.6 and greater than for
the volume terms, but it is small. For  $M^{*}_{\infty}/M$  = 0.7   it is not possible
to achieve the same spin-orbit potential as   for the $M^{*}_{\infty}/M$  = 0.6
case. To obtain this effect a tensor coupling of the vector field has to be added
                          \cite{Estal}.

   \section{Conclusions.}
   We have studied the influence on   the surface properties of the new terms arising from
an energy functional based on relativistic effective field theory. 
\pagebreak
 \begin{figure}[h!]
   \caption{
                                                (a)        Contours of constant
  $E\dr{s}$ (solid line) and $t$ (dashed line) in the  $\alpha_1-\alpha_2$ plane.
  (b) Spin-orbit potential for the cases indicated in the legend.}
         \vspace{3mm}
         \setlength{\unitlength}{1mm}
         \begin{picture}(80,80)
         \put(26,03){\epsfxsize=6.00cm  \epsfbox{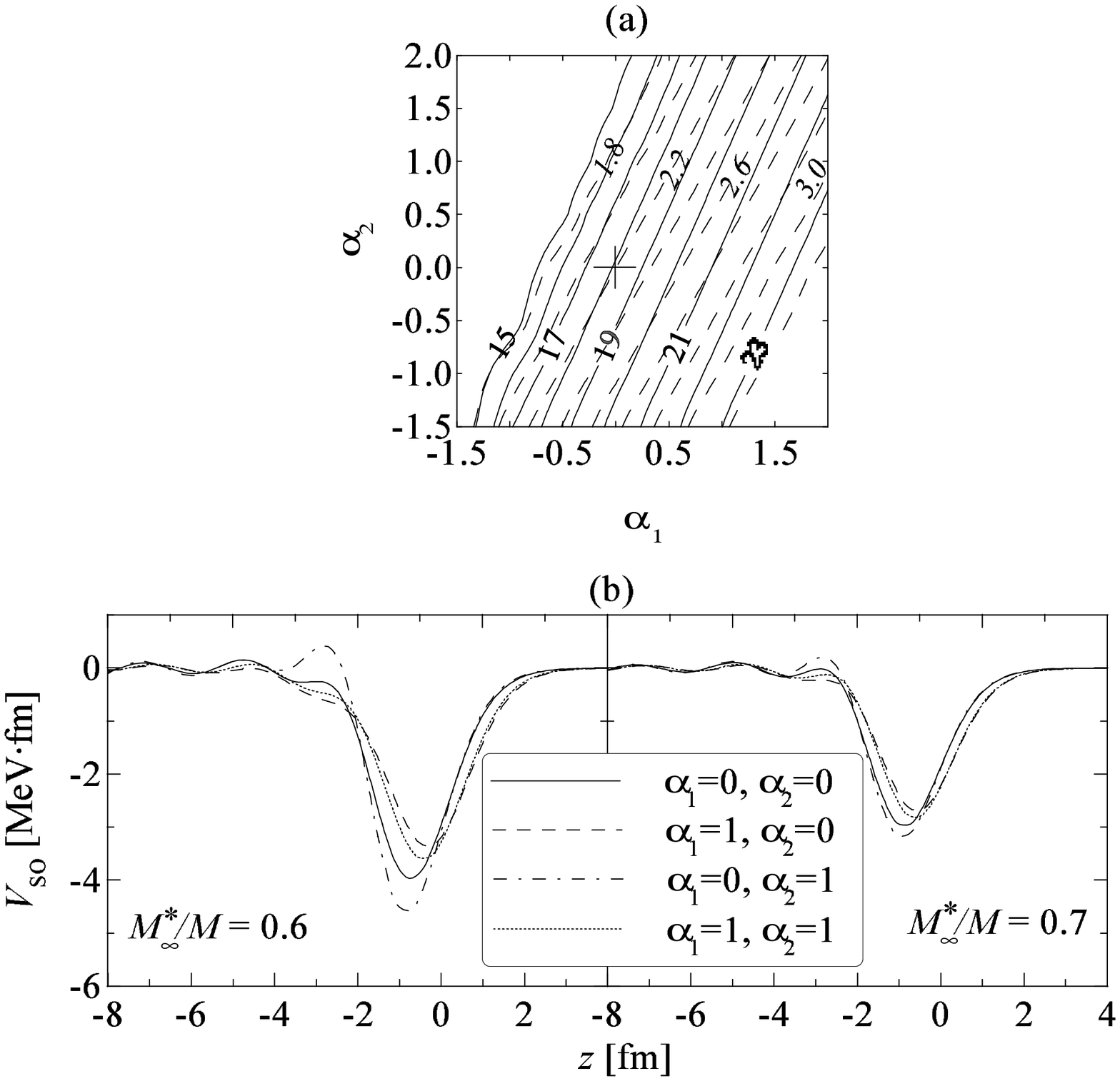}}
         \end{picture}
         \end{figure}
\vspace{-1.30cm}

                       The volume terms (quartic
vector and scalar-vector self-interactions) show a small influence on the
surface properties. Nevertheless, the introduction of terms proportional
to the gradients of the meson fields can be used to improve the surface properties
of a given parametrization taking into account that these terms do not play any role
on the saturation properties. Concerning the spin-orbit potential, the influence
of all these terms is small. The results of this investigation as well as the study
of the properties of asymmetric nuclear matter will be presented
                                 \mbox{elsewhere \cite{Estal}.}

\end{document}